
\input phyzzx
\sequentialequations
\overfullrule=0pt
\tolerance=5000
\nopubblock
\twelvepoint
\def\Q{Q$^2$\ }

\line{\hfill IASSNS-HEP-93/69}
\line{\hfill August 1993}
\line{\hfill hep-ph/9311302}
\titlepage
\title{Status of QCD\foot{ Invited talk given at
the Cornell Lepton-Photon  conference,
Ithaca N. Y. August 1993.  This manuscript closely follows the
original talk.}}
\author{Frank Wilczek\foot{Research supported in part by DOE
grant
DE-FG02-90ER40542}}
\vskip.2cm
\centerline{{\it School of Natural Sciences}}
\centerline{{\it Institute for Advanced Study}}
\centerline{{\it Olden Lane}}
\centerline{{\it Princeton, N.J. 08540}}
\bigskip
\bigskip
\bigskip

\REF\pick{A. Pickering, ``Constructing Quarks,"
University of Chicago Press, (Chicago, 1984).}

\REF\bethke{S. Bethke, S., Proceedings of the XXVI International
Conference on High Energy Physics, Dallas, TX 1992.
  The Figures as presented here are revised and
updated from the Dallas talk.  I thank Professor Bethke for
making these Figures available to me.}

\REF\shapiro{M. Shapiro, these proceedings.}

\REF\alt{G. Altarelli, in ``Twenty Years of QCD," ed. P. Zerwas, H.
Kastrup (2 vols.), World Scientific, (Singapore, 1993).  These volumes
are an excellent source for useful, up-to-date reference material on
QCD.}

\REF\ellis{J. Ellis, S. Kelley, D. V. Nanopoulos, Phys. Lett. {\bf B260}
(1991) 131; U. Amaldi, W. de Boer, H. F\"urstenau, Phys. Lett. {\bf
B260} (1991) 447; U. Amaldi, et al., Phys. Lett. {\bf B281} (1992) 374;
P. Langacker, M. Luo, Phys. Rev. {\bf D44} (1991) 817; J. Ellis, S.
Kelley, D. V. Nanopoulos, Nucl. Phys. {\bf B373} (1992) 55.  For
up-to-date information see W. de Boer, R. Ehret, D. Kazakov,
``Constraints on SUSY Masses in Supersymmetric Grand Unified Theories,"
Karlsruhe preprint, hep-ph/9308238.}

\REF\braat{E. Braaten, S. Narison, A. Pich, Nucl. Phys. {\bf B373}
(1992) 581.}

\REF\muel{A. Mueller, paper in Ref. [4].}

\REF\zj{For a good review of this work up to 1980, with full references,
see J. Zinn-Justin, Phys. Rep. {\bf 70} (1981) 109.  More recent
developments can be traced from the references in Mueller, Ref. [7].}

\REF\aleph{ALEPH Group, Phys. Lett. {\bf B307} (1993) 209.}

\REF\elkh{A pioneering effort along these lines is A. El-Khadra, et al.,
Phys. Rev. Lett. {\bf 69} (1992) 729.}

\REF\mac{P. Mackenzie, these proceedings.}

\REF\ccfr{Data from the CCFR collaboration.}

\REF\gw{D. Gross, F. Wilczek, Phys. Rev. {\bf D8} (1973) 3633, {\bf D9}
(1974) 980; H. Georgi, H. D. Politzer, Phys. Rev. {\bf D9} (1974) 416.}

\REF\sld{Data from the SLD collaboration.}

\REF\ua{Data from the UA1 collaboration.}

\REF\bern{For an up-to-date review see Z. Bern, ``String-Based
Perturbative Methods for
Gauge Theories," TASI Lectures 1992.}

\REF\bdk{Z. Bern, L. Dixon, D. Kosower, Phys. Rev. Lett. {\bf B70}
(1993) 2667.}

\REF\bd{Z. Bern,  D. C. Dunbar, and T. Shimada,  Phys. Lett. {\bf B312},
(1993), 207.}

\REF\parke{A superb summary of the pre-string work is given in
M. Mangano, S. Parke, Physics Reports {\bf 200} (1991) 301.}

\REF\lam{The most complete field-theoretic discussion containing
important new ideas is  C. S. Lam, "Spinor Helicity Technique and String
Reorganizatin for Multiloop Diagrams", McGill preprint (June 1993),
hep-ph/9308289.}

\REF\levin{For a review see E. Levin, in Ref. [4].}

\REF\bal{I. Balitsky, V. Braun, Phys. Lett. {\bf B314} (1993) 237.}

\REF\webb{For reviews with extensive references see the papers of
B. Webber and J. Drees in Ref. [4].}

\REF\bur{For a review with extensive references see the paper of A.
Buras in Ref. [4].}

\REF\wise{M. Wise, these Proceedings.}

\REF\bohr{In the sense of Bohr.  According to Bohr
a simple truth is a statement whose opposite is a falsehood;
a profound truth is a statement
whose opposite is also a profound truth.}

\REF\stat{For recent work see, K. Rajagopal, F. Wilczek,
Nucl. Phys. {\bf B399} (1993) 395;  These papers contain extensive
references to the earlier work.}

\REF\brown{F. Brown, et al., Phys. Rev. Lett. {\bf 65} (1990) 2491;
C. Bernard, et al., Phys. Rev. {\bf D45} (1992) 3854; S. Gottlieb,
et al., Phys. Rev. {\bf D47} (1993) 3619.}

\REF\actual{Since the actual talk a beautiful theoretical analysis
of the numerial data has appeared, F. Karsch, ``Scaling of
Pseudo-Critical Couplings in Two-Flavor QCD," HLRZ Preprint 93-62 (1993),
hep-lat/9309022.
  The situation may already
be substantially better than I thought in August.}

\REF\kap{D. Kaplan, A. Nelson, Phys. Lett. {\bf B175} (1986) 57.}

\REF\pol{H. D. Politzer, M. Wise, Phys. Lett. {\bf B273} (1991) 156.}

\REF\bb{G. Brown, H. Bethe, ``A Scenario for a Large Number of Low
Mass Black Holes in the Galaxy," Caltech Preprint, (1993).}

\REF\helf{D. Helfland, R. Becker, Nature {\bf 307} (1984) 215.}

\REF\wit{E. Witten, Phys. Rev. {\bf D20} (1984) 272; A Mann, H.
Primakoff, Phys. Rev. {\bf D22} (1980) 1115; A. Bodner, Phys. Rev.
{\bf D4} (1971) 1601; S. Chin, A Kerman, Phys. Rev. Lett. {\bf 43}
(1978) 1291.}

\REF\far{E. Farhi, R. Jaffe, Phys. Rev. {\bf D30} (1984) 2379;
M. Berger, R. Jaffe, Phys. Rev. {\bf C35} (1987) 213; E. Gibson,
R. Jaffe, Phys. Rev. Lett. {\bf 76} (1993) 332.  There is a book
on the subject: ``Strange Quark Matter in Physics and Astrophysics,"
ed. J. Madsen, P. Haensel, Nucl. Phys. (Proc. Supp.) {\bf B24} (1991).}

\REF\schaf{J. Schaffner, C. Dover, A. Gal, C. Greiner, H. St\"ocker,
Phys. Rev. Lett. {\bf 71} (1993) 1328, and references therein.}

\REF\mul{An excellent overview, with extensive references, is B.
Muller, ``Physics of the Quark-Gluon Plasma," Duke Preprint TH-92-36
(1992).}

\REF\van{L. Van Hove, Phys. Lett. {\bf 118B} (1982) 138; Z. Phys.
{\bf C21} (1983) 93.}

\REF\raja{A recent work is K. Rajagopal, F. Wilczek, Nucl. Phys.
{\bf B404} (1993) 577.  This contains extensive references to the
earlier literature.  Pioneers of the subject include Anselm, Ryskin,
Blaizot, Krzywicki, Bjorken, Taylor and Kowalski.}

\REF\call{C. Callan, R. Dashen, D. Gross, Phys. Lett. {\bf 63B} (1976)
334; R. Jackiw, C. Rebbi, Phys. Rev. Lett. {\bf 37} (1976) 172.}

\REF\nara{R. Narayanan, H. Neuberger, Phys. Lett. {\bf B302} (1993) 62;
D. Kaplan, Phys. Lett. {\bf B288} (1992) 3421.}

\REF\strassler{M. Strassler, {\it Nucl. Phys.\/} {\bf B385}
(1992) 145.}

\REF\shatashvili{In approaching this question, it is probably important
to be aware of the work of A. Polyakov, Mod. Phys. Lett. {\bf A3}
(1988) 455, who shows how to represent spin as a flow, and
its generalizations in A. Alekseev, S. Shatashvili, Mod. Phys. Lett.
{\bf A3} (1988) 1551; A. Alekseev, L. Fadeev, S. Shatashvili, Jour.
of Geometry and Physics {\bf 5} (1989) 391.}

\REF\deu{See especially, D. Deutsch, Proc. R. Acad. London {\bf A400}
(1985) 97.}

\REF\holz{C. Holzhey, Princeton University Thesis (unpublished, 1993),
reported by F. Wilczek in ``Black Holes, Membranes, Wormholes and
Superstrings," ed. S. Kalara, D. Nanopoulos, World Scientific, Singapore
(1993).}

\REF\soe{M. Srednicki, Phys. Rev. Lett. {\bf 71} (1993) 666.}

\REF\suss{See in this regard, L. Susskind, Phys. Rev. Lett. {\bf 71}
(1993) 2367.}

\REF\lars{I have benefitted from extensive ongoing discussions with
F. Larson, C. Nayak and C. Callan on these subjects.}

\voffset=.3in
\hoffset=.2in
\singlespace

I have been asked to discuss the status of QCD.
It seems to me
that there are three main points to be made about the
present status of
QCD:

$\bullet$ QCD is right, and we can do many beautiful things with it.

$\bullet$ There are several important concrete problems that lie just
beyond the edge of our current understanding.

$\bullet$ There are some foundational issues in QCD, and some recent
developments, that may point toward entirely new directions.

These points will, I believe,
emerge quite clearly from the following
more detailed discussion.
The discussion will be in three parts.
I'll first discuss elementary
processes, then more complicated processes, and then finally
foundational issues.

\chapter{Elementary Processes}

\section{Testing QCD, measuring $\alpha_s$}

Clearly the best way to test QCD, which is formulated as
a microscopic theory, is to test that the elementary processes it
postulates in fact occur and are correctly described by the theory.
Even verifying
the first part of this -- that the processes occur -- is
highly non-trivial from a logical point of view.  Indeed
the entities
in terms of which QCD is formulated, quarks and gluons, cannot be
studied at leisure in isolation.  By now however all
practicing physicists, if
not all philosophers of physics [\pick ], are comfortable
with the very tangible reality of quarks and gluons.  If
theorists had not invented them beforehand, quarks and gluons
would surely
have
been ``discovered'' as the only adequate description of hadronic
jets in high-energy collisions.

Because of asymptotic freedom, perturbative
QCD supplies quantitative
predictions for a wide variety of experiments.
I think it is fair to say that no major
discrepancy between theory
and experiment has emerged thus far.
The question naturally arises, just how stringent are the
tests?
A popular quantitative measure of this
is how tightly the strong coupling constant is constrained.
This is useful also in checking the mutual consistency of
the QCD fits made in (often {\it very}) different experiments, and
in testing models of unification.

This subject, experimental
determination of the strong coupling,
has been thoroughly reviewed recently
at the Dallas Conference by Bethke [\bethke ] and
this morning by Marjorie Shapiro [\shapiro ].
Therefore I will not attempt to do
justice to the many experiments in detail, but rather confine myself
to an
impressionistic survey.

Because of the way that the
strong coupling constant runs,
that is decreasing as \Q increases,
testing QCD by measuring the strong coupling constant as a function of
\Q has a two-faced character.  At large \Q the predictions are
precise -- the coupling gets small,
and we can do accurate calculations.
Also mass corrections, which are non-perturbative and
generally difficult to
estimate theoretically, are
suppressed by powers of mass$^2$ over Q$^2$.  On the other hand
such large \Q measurements
are limited in their power to resolve
possible values of $\alpha_s$
quantitatively.
At
small \Q the theory is much harder to control and make
precise, but if you are interested in quantitative results for
$\alpha_s$ there is a large premium for working at small \Q .

These features are
clearly evident in the classic plot of coupling constants
measured at different \Q  [Figure 1].
You see that at low \Q there is a big spread in the
couplings, so that is where
you can determine what $\alpha_s$
or equivalently $\Lambda_{\rm QCD}$
is.  On the other hand for very large \Q we get -- remarkably --
a more or
less unique prediction for the value of the physical parameter
$\alpha_s$ (Q$^2$).  Almost any
reasonable value of $\Lambda_{\rm QCD}$, that is
anything that is sensible to regard as an overall scale for strong
interactions,
will give you within about
10\% the same results for $\alpha_s$ at
large Q$^2$.

If your goal is simply to check that
QCD is right, then you want a unique prediction
and the high energy
processes are particularly favorable.   But
if you want to be quantitative about $\alpha_s$, then
the low energy determinations have a big advantage.
This is the two-faced character I mentioned.


Figures 1, 2 and 3, taken from Bethke (with a small addition, to be
discussed presently), summarize the quantitative results from a wide
variety of experiments which measure $\alpha_s$.  Figure 1
shows the range of theoretical expectations and the relevant
experiments in each \Q range, Figure 2 summarizes their
theoretical foundations
and uncertainties, and Figure 3 exhibits the results in tabular
form.  In these Figures DIS means
deep inelastic scattering, NLO means next leading order, NNLO means
next to next to leading order, and MC means Monte Carlo.   Finally
GLS refers to Gross-Llewlyn-Smith.  The GLS determination of
$\alpha_s$ is based on the difference between the parton
model prediction for their sum rule, which highlights the baryon number
content of the nucleon, and the measured value.
The overall consistency of the results, extracted
as they are from widely different
experiments covering a vast kinematic range,
is certainly most impressive.

Most of what is in these Figures has been discussed by Bethke and
by Shapiro, and also by Altarelli [\alt ],
with greater insight and authority than I can bring to the subject.
Anyone really interested in the quality of the fits should
turn to their reviews.  I would, however, like to add
a few supplementary
comments.

First I would like to call attention to the
final entries on the right of Figure 1 and the bottom of Figures
2 and 3.
These refer to a determination
of the strong coupling which is different in character from the more direct
QCD measurements [\ellis ].  Its inputs are $\alpha_{\rm QED}$,
that is the fine structure
constant, and the  Weinberg angle.  In the context of grand
unification, the three gauge coupling constants of the standard model
are not independent, so that -- for a given model of
unification -- any two determine the third.  These considerations
give determinations of the strong coupling whose {\it precision\/}
is comparable to that from the other, more conventional determinations.
The {\it accuracy\/} of the determination
is another matter.  It depends on the specific
model of unification, as we shall see.

The uncertainties in this determination of $\alpha_s$ are rather
different from those in the other cases.
Mundane problems like hadronization corrections, non-perturbative
contributions, higher twists, small lattice sizes and the quenched
approximation -- none of these is important.  The unification of
couplings is the ultimate large \Q process, and sheds the dross of
low-energy QCD.  However there remains
a very important {\it caveat\/}
to these determinations, namely that they are ambiguous
and could
be completely wrong.
There is precious little independent evidence for the unification
theories on which they are based.  They can't all be right, because
you get slightly
different determinations depending on what you assume about
unification.

In fact the minimal grand unified
model, based on SU(5), leads to a
small but clear-cut conflict with the data.
Given the enormity of the extrapolations involved it is
most remarkable
that the prediction works out even as well as it does.  It might
be wise to stop at this point, declaring victory for the principles
of unification and of quantum field theory extended down to distances
of 10$^{15}~$GeV while acknowledging that there's some
fuzziness in
the details.
On the other hand, quite remarkably,
the prediction following from SU(5) unification
of the
minimal supersymmetric
extension of the standard model
is in adequate agreement with the data [\ellis].
It is certainly tempting to take this agreement
as an
indication that
simple ideas about
unification and low-energy
supersymmetry have something to do with reality.


Another entry in
the table of Figure 3
is particularly interesting and provocative, and suggests
some later developments, so I want to go into a little more detail
regarding it.
It is the determination of $\alpha_s$ from
QCD with
corrections to the tau lepton lifetime [\braat].
Tau lepton decay of course is a very low
energy process by the standards of LEP or most other QCD tests.
So we can expect, in line with the previous discussion,
that the prediction will perhaps
be delicate but on the other hand it will have
a favorable
lever arm for determining $\alpha_s$.

The quantity governing inclusive $\tau$ decay, the square of
the amplitude summed over intermediate states, is
the vacuum matrix element of a current product.  The product
could be the vector
current times the vector current, or axial current times axial, or
axial times vector, with different isospins and in principle
different strangeness -- all these can be separated
out by projecting on the quantum numbers of the final state.
Suppressing all indices,
the relevant object for theoretical analysis is
the vacuum matrix
element of the current product
$$
J(Q)J(-Q) \approx {\cal C}_{\rm I} (Q^2 ) {\rm I} (0) ~+~
  {\cal C}_{\bar \psi \psi} (Q^2) m \bar \psi \psi (0) ~+~
  {\cal C}_{\rm GG}(Q^2)  {\rm tr} {\rm G}_{\mu\nu} {\rm G}_{\mu\nu} (0)
   ~+~ ... ~.
$$
The current product in field theory is
evaluated, as above, using
the operative product expansion.
If we had a complete set of operators on the right hand side we'd
expect to get a complete representation of this current product.  The
successive operators have higher and higher mass dimension and
therefore their coefficients, since the overall dimension has to be
the same, have larger and larger powers of inverse \Q.  So, given that
their matrix elements are characterized by some typical strong
interaction scale we'll be getting higher and higher
powers of the strong interactions
scale over Q$^2$.  Keeping the first few terms
should  be a good approximation even at 1.8 GeV.  It is very helpful
that the mass dimensions of the gauge invariant operators start at
4.

The Wilson
coefficients, the operator product coefficients ${\cal C}$ above,
obey renormalization
group equations.  They can be calculated in perturbation
theory in the effective coupling at large Q$^2$,
of course.
However, at \Q of approximately $m_\tau^2$
we cannot simply ignore plausible
non-perturbative corrections and still guarantee worthwhile accuracy.
A term of the form $\Lambda_{\rm QCD}^2 / Q^2$ would show up,
through the mechanism of dimensional transmutation, as a contribution
proportional to $\exp {(-c/\alpha_s) }$ in this coefficient,
where $c$ is a calculable numerical constant.
It is an important question whether there is such a contribution,
because if there were, and they were not under tight control, it is
formally of such a magnitude as to ruin the useful precision of the
predictions.
Such a correction would be bigger than the ones coming from
higher operators
because these operators have dimension 4, so
their coefficients have \Q
over $\Lambda^2$ squared, which is {\it a priori\/} smaller.

Mueller [\muel] has given an important,
although not entirely rigorous, argument
that no  $\Lambda^2/Q^2$ term  can appear.
The argument is a little
technical, so I won't be able to do it full justice here
but I will attempt to convey
the main idea.
The argument is based on the idea that at each successive
power of 1 over \Q
one can make the perturbation series in QCD, which is
a badly divergent series in general, at least almost
convergent, that is Borel summable,
by removing a finite number of obstructions.  Furthermore the
obstructions
are captured and parameterized by the low dimension operators
mentioned before.  Once these obstructions are removed, the
remaining (processed) perturbation expression converges on the
correct result for the full theory.  Neither in the
obstructions nor in the residual perturbative expression do the
potentially dangerous terms occur -- which means
that they don't occur at all.

Maybe I should draw a picture of this [Figure 4].
One has the current product, and one is doing an analysis of
its behavior when large virtual momentum is flowing through the
current lines.
The principle of the operator product expansion is
to exhibit the powers of \Q by
breaking the propagators in the graph into hard and soft parts.
Any soft part costs you a power of one over \Q so you want the
minimal number.
If
you just take out a couple of lines
you have one of those low dimension
operators, so those are interpreted as the
operators,
with the understanding that
now all internal propagators in the graph are hard.
This really provides the rigorous
{\it definition\/} of these operators, which is non-trivial to do
and inevitably
introduces scale dependence due to ultraviolet divergences in
products of the basic field operators at the same point.

The Wilson coefficients in the operator product are computed from
these graphs with hard internal propagators.  They obey renormalization
group equations, which tell you how they change if the nominal
scale dividing ``soft'' from ``hard'' is re-defined.  These equations
lead by the usual arguments to the appearance of the running coupling
constant in the graphical evaluation of the Wilson coefficient.
Formally, then, because of asymptotic freedom, one can evaluate this
coefficient perturbatively.  In this evaluation, to all orders of
perturbation theory the only scale dependence comes from the
dependence of the running coupling on Q$^2$, and it is logarithmic
in Q$^2$.
You have to show, to complete the
argument that eliminated the possibility of
power law corrections in perturbation theory,
that they don't arise in the true answer.
This is not obvious, because the perturbation theory
does not converge.  Thus
it is overly optimistic to suppose
that it converges on the true answer.

There has been a lot of theoretical work, starting with
Dyson, on the question why perturbation
theory diverges, and how in favorable cases one can salvage
valid results from it nonetheless.
Although a technical account would be out of place here, let me
give you a little taste of how it goes  [\zj ].
One starts with a formal perturbation series
$$
f(\alpha )~\sim~ \Sigma_n~c_n~(\alpha )^n
$$
that does not converge, because the $c_n$ do not fall off rapidly
enough with $n$.  It may happen however that the modified series
$$
g(\alpha )~=~ \Sigma_n~ ({c_n\over n! }) (\alpha )^n
$$
does converge and defines a legitimate function $g(z)$.  If so,
then
$$
\tilde f (\alpha ) ~\equiv~ \int_0^\infty~ ds~ s^{-1}
  ~e^{-s/\alpha }~g(s)~,
$$
if it exists, defines a legitimate function whose formal power series
expansion agrees with that of f.  The underlying point is that a normal
power series is limited in its radius of convergence by the nearest
singularity, whereas a more flexible representation may converge in
a different shaped region.  In quantum field theory there is generally
a singularity at arbitrarily small negative $\alpha$; but in favorable
cases the Borel transform exists and its inversion
uniquely defines a function
analytic in a wedge excluding the negative real axis.  This function
then gives strict meaning to the divergent perturbation series.
One can show that when it works this procedure defines amplitudes which
satisfy the axioms of quantum field theory such as causality and
unitarity.  The usual demonstrations
that these properties hold order by order in perturbation theory
can be adapted
to the re-processed version, which is more complicated but has the
virtue of actually defining an answer.  In fact we can agree
that it gives {\it the\/}
answer, since after all the whole point of quantum field theory is
to give non-trivial realizations of the axioms, and that is what we
have found.

QCD is not quite so favorable as this ideal, which
occurs only for massive super-renormalizable theories
in low dimensions. There are several known obstructions
to Borel summability in QCD,
which go by frightening names: ultraviolet
and infrared renormalons, instantons,
and threshold-induced oscillations.
What Mueller did was to analyze these known sources of
possible dangerous terms.  He argued that the infrared renormalons
are essentially just the higher-order terms in the operator product
expansion, the ultraviolet renormalons generate singularities
in $g(\alpha)$ away from
the real axis whose influence on the truncated form
of $g(\alpha)$ one actually computes can be minimized by judicious
mappings in the $\alpha$ plane, that the threshold-induced
oscillations are negligible quantitatively, and that the
instanton contribution is both small and in principle calculable.

So now I have fleshed out my earlier description of Mueller's
argument a bit.  The key underlying assumption is that the known
obstacles to Borel summability are the only ones.  In principle,
one can test this circle of ideas
by calculating the operator product coefficients
directly in the full theory (\ie \ numerically, using lattice gauge
theory techniques).  If they were to
fail, it would signify that there is an
important gap in our understanding of quantum field theory.



On the experimental side,
the Aleph
group has tested the framework leading to
this operator product expansion by comparing
the resulting specific predictions
for decay into semi-inclusive final states
with specific quantum numbers,
including the \Q dependence (which you can look at by looking at
final states of different mass) [\aleph ].  They got a good fit with no one
over
\Q term and with matrix elements of
the lowest dimension relevant operators  $m \bar \psi \psi$,
${\rm tr} {\rm G}_{\mu\nu} {\rm G}_{\mu\nu}$
fitted to other experiments.  These quantities also appear in
other similar applications, where observed hadron parameters
are correlated
using the so-called QCD or ITEP sum
rules, which arise by saturated various operator products.
By taking suitable moments one can define quantities that are
insensitive to the higher dimension operators, and for these the
predictions of perturbative QCD are especially stringent.

I went into some detail into the analysis of tau decay
because I think it's not only important in itself but
quite fundamental, and it connects with many other issues.
In particular,
this kind of argument could potentially provide a rigorous foundation
for the QCD or ITEP sum rules which are the basis of a very successful
phenomenology.  Also similar analyses could be attempted in relatively
low \Q e$^+$e$^-$ annihilation and possibly in
deep inelastic scattering.

As far as the specific issue
of determining $\alpha_s$ is concerned,
however, all this will probably soon be moot.
The character of the enterprise of
determining $\alpha_s$ and testing QCD quantitatively is,
as we speak,
undergoing a
decisive transformation.  It is becoming obvious that
before very long direct simulation of heavy quark systems using
the full non-perturbative force of QCD, simply by discretizing the
theory and doing the integrals numerically (lattice QCD), will
provide the best determinations.
In this application, of course, one works at reasonably
small \Q and includes
the whole theory.  Also there are directly measurable effects
such as hyperfine splittings which are directly proportional to
a power of $\alpha_s$, so that instead of looking for a small correction
to parton behavior one is getting
the strong coupling parameter out front.  When we consider
these
favorable theoretical
features together
with the precision possible in the corresponding
measurements, it becomes evident
that this method is the future of $\alpha_s$
determination.
Indeed, as we heard from Paul Mackenzie,
it may be that the future is now [10,11].

So much for determinations of $\alpha_s$.
I'd like to emphasize something
that may get lost in the enthusiasm for presenting numbers
for $\alpha_s(M_{\rm W})$, and
competing to see who has the most exact.  That is, that numerical
accuracy in determining $\alpha_s$ is not a completely adequate
figure of
merit for experimental tests of QCD.
Let us suppose, what I think is now eminently reasonable to suppose,
that QCD correctly describes many processes in nature -- that the
quarks, gluons, and Yang-Mills dynamics are here to stay.  Then in
``testing'' QCD perhaps the point of greatest interest is not to
continue to verify yet again that these degrees of freedom exist
and their mutual interactions are correctly described, but to test
interesting hypotheses for new physics.  Let me give an example.
Yesterday when I
was talking with one of my colleagues in preparing this talk, we
both had a laugh over how silly it was that people were testing the
flavor independence of QCD, checking that their determinations of
$\alpha_s$ in different flavor channels were consistent.
Then I thought a little more about it,
as undoubtedly the experimentalists had
beforehand, and realized that it's not
silly at all.  For instance the existence of
a Higgs particle would be very interesting.
The coupling of a Higgs particle to quarks is
typically to the quark mass, and
diagonal in flavor.   It would therefore show
up mostly as a quantitative correction to heavy quark processes otherwise
dominated by QCD.  If
the Higgs particle had vacuum quantum numbers and
complicated decay channels events involving its exchange (or production)
might not be trivial to distinguish from normal QCD events.
However
presumably one way it would show up would be as a
correction to flavor independence of the nominal $\alpha_s$.

The moral of this little tale
is that qualitative tests of salient features of QCD are neither
silly
nor obsolete -- and
besides they're a lot of fun.
So I'd like to spend a few
minutes touring a few of these.

Figure 5 [\ccfr ] is historic because it shows the verification
very first type of quantitative prediction to be extracted
from QCD [\gw ].
Little did we
dream that it would ever be measured so accurately, but here it is,
the \Q dependence of the non-singlet F$_3$ structure function.
This is a
particularly clean application of QCD because it comes from the
operator product expansion, the corrections are under good
control, and the comparatively poorly known gluon distributions
do not enter.
You see that the
structure function does just what it's supposed to do, dropping
steeply with \Q at large
x and less steeply  at small x.


A pretty test, especially
satisfying because it is intuitively clear
that
you are seeing the  running of the coupling constant in a single process,
is exhibited in Figure 6 [\sld ].  It
is the three-jet production rate as a function of energy in
e$^+$e$^-$ annihilation (at a
fixed value of the Durham cutoff [\bethke ]).
This is, of course, a direct measure of the probability of radiating
a gluon and thus of its effective coupling.
Plotted as a function of energy presented versus an inverse
logarithm the rate should lie to first approximation on
a straight line,
and you see that it works very nicely over a wide range of energies,
with the inverse coupling itself changing by about 50\%.
This was all done originally for e+e- annihilation; at the
Conference we have a report from Fermilab
experiment 665 where a similar
phenomenon is now observed in deep inelastic scattering.

Direct evidence for
the Rutherford cross-section in jet physics,
showing that indeed the dominant hard process
in the strong interaction is the exchange of elementary  spin one
boson, is shown in Figures 7 and 8 [\ua ].
It is worthwhile to emphasize that at very high energy,
I think at upgraded Fermilab and certainly at LHC, in the bulk
of hard events
one will be seeing two gluons scattering into
two gluons.  Thus the triple gluon vertex characteristic
of the nonabelian gauge interaction will be directly responsible
for most of what
goes on there.

So much for showing that QCD is right.
I think there is no reasonable doubt
anymore that it is right, in the strongest sense that a
scientific theory can be right.  That is, QCD will
be used for the forseeable
future, in essentially its present form, as the primary
description of a wide body
of phenomena.

\section{New methods in QCD perturbation theory}

In searching for new physics at high energies, possibly
involving new
kinds of heavy particles
that decay into several light particles and several jets,
you can have severe QCD backgrounds.  It is therefore quite
important to be able to compute complicated processes in QCD. Such
computations can get out of hand quite rapidly due
to the proliferation
of Feynman graphs, each of which corresponds to an amplitude
with many terms due to the complicated vertices.

Fortunately, lately remarkable
improvements in the technique of calculating
complicated processes at high energy in  perturbation theory
have been discovered.
Several
insights, including important ones from string theory,
have been combined into an approach
powerful enough to address processes
that would have seemed hopelessly
out of reach even a year ago [\bern ].  Perhaps
the most stunning example appears in a recent Physical Review Letter
by Bern, Dixon and Kosower [\bdk ],
where they calculated the rate for two
gluons goes to three gluons to one loop order in QCD.
(Although it is somewhat off the immediate point, I would
like  to mention that Bern and Dunbar [\bd ] have also computed
graviton-graviton processes scatterings in a  compact form.
If you've ever attempted to calculate
even very simple graviton processes in perturbation theory, you'll
be suitably impressed.)

The  techniques
used to do these calculations are of great interest in themselves.
They are techniques that arise naturally in string theory,
although some of them could have been, and to some extent were,
developed independently in
other contexts [19,20].

I'll say a bit more about this subject toward the end.


\section{The boundaries of perturbative QCD}

Lately there have
been very pretty and promising developments pushing the
edge of perturbation theory, pushing into domains where you can just
barely justify the use of perturbation theory, or
where it's starting to break
down.

A well known phenomenon that follows from the evolution of the
structure functions is that density of wee gluons, as measured in the
gluon structure function,
piles up at small x as \Q grows.
This arises because
if you look at higher \Q it has an admixture of two
gluons with smaller \Q and smaller x, because the Hamiltonian is not
diagonal in the states of definite gluon number, but connects these
configurations.
If in the evolution the number of
gluons piles high enough eventually you
have to take into account not
only that fragmentation process
but also the inverse process: two
of the existing gluons already in the nucleon,
combine into one gluon.  This is a qualitatively distinct
and quite interesting
process from the fragmentation,
because it  reflects the properties of the nucleon
-- the nucleon can no longer be regarded as just a container for a dilute gas
of quarks and glue, even at large \Q .

This effect is estimated to be
quantitatively small until you get to
very small x, but it may turn out to become
very important as you get to
exceedingly low x as is being probed at HERA.  Presumably
balancing the fragmentation  against the recombination
leads to a saturated gluon density which
is proportional to one over Q$^2$ times one over some
radius$^2$ over which
the gluons are spread.  (This calculation should
really be done in the infinite
momentum frame, so there are only 2 transverse directions.)
The fragmentation and recombination are of the same order, so
it's a non-perturbative problem to find
the balance.  Fortunately however
there is still a small expansion parameter one over
Q$^2$R$^2$, and progress can be made [\levin ].
I'm giving you a cartoon introduction to the theory, which is actually
very technical and difficult to understand in its present state.

Another way of
thinking about the main phenomena is to think of it as shadowing,
that is as the effect that gluons
as they are attempting propagate through the nucleus can be
absorbed by other gluons.  But here, unlike in any other situation
in physics I can think of, we have a case of significant shadowing by
simple fundamental objects, namely the gluons, and at small coupling.
Thus it is an interesting challenge
actually to calculate it quantitatively.
This is I think a very promising
area at the edge of perturbation theory, where fundamental ideas about
diffractive scattering can be tested in a precise way.

Another related
possibility I find very intriguing is suggested
in a recent paper by Balitsky and Braun [\bal ].  Their idea
is that in this small x
region where you have lots of soft gluons they can
gang up, so to speak, and produce instantons.   Balitsky and
Braun estimate in fact
that the instanton contribution to the F$_2$ structure function
is 2 to 5\% at \Q=400 GeV$^2$.
That's quantitatively small as a fraction of the cross-section,
but the events involved would be quite
striking, they would have a high multiplicity proportional to
$2\pi\over \alpha_s$
in one rapidity unit with large transverse momentum.
That's what you would find accompanying the current jet.
This
suggestion
of Balitsky and Braun
definitely needs more work to make it quantitative and
precise but I think it is extremely promising.  It is closely related
to problems people discuss in the electroweak interactions, where it
has even been suggested that the electroweak interaction becomes strong
due to contributions from
instantons at high energies.  I think
for the weak interaction this is a very dubious proposition, but
in the Balitsky-Braun process you have a version of the same ideas
which might be sensible and accessible to experiment.
Certainly we can look forward to learning
interesting things about quantum field theory by being forced to address,
under the prod of ongoing experiments,
challenging problems at the frontier of perturbation theory.

Finally, on a different
edge of perturbation theory, I'd like to show you one
little tip of a very large iceberg [\webb ], a simple representative coherence
effect, just to show the inadequacy of overly naive partonology.
(See Figure 9).
This is work by L3 collaboration; I
think it's quite pretty.  The question is: do jets fragment
independently?
They do according to a parton model; that's the core assumption
of the model.
What's measured to test this is
what happens if you have a three jet event by an
appropriate criterion which cuts on the invariant mass in the
jets, and then make the
resolution finer, so you're demanding that the jets
have smaller invariant
mass.  What is the probability that these jets, when looked at finer,
fragment?

Now if the jets fragmented independently
you would get a very large probability from the three jet events
because the gluon has a large color charge and just adds to the other
two fragmentation probabilities for just the quark and antiquark.
Thus you'd have a large
contribution which simply adds to the
two-jet
contribution.  But in QCD these things are not
independent, there is interference between the different processes
because they are described by quantum-mechanical amplitudes, not
probabilities.
The accurate calculation shows that the fragmentation probability,
which according to
the incoherent model goes up,
instead should actually, according to QCD,
go down.
The accurate calculation agrees very nicely with the
data.

Thus we have both quantitative and qualitative tests of the
elementary processes of QCD, and also ``elementary''
processes that are at the hairy edge of existing quantitative methods.

\chapter{More Complicated Processes}

\section{QCD as a service subject}

Given that QCD is right,
it ought to become a
service subject for many other parts of physics and perhaps ultimately
for natural philosophy.

Here is a little table [Figure 10] summarizing some of the services
we should expect QCD to provide for
fields of physics.  In every case there is much room for
improvement, and these fields of ``applied QCD'' present
many intellectual challenges.

$\bullet$ {\it For high-energy physics}: It may seem peculiar
to speak of the application of QCD to high-energy physics, since
QCD is usually considered a part of high-energy physics.  However
``today's sensation is tomorrow's calibration'', and if one defines
high energy physics by its frontier, by the search for new fundamental
laws, then one may properly speak of QCD as a service subject for
high-energy physics.

As I mentioned before, experimentalists doing experiments at the
highest energies rely heavily on QCD to estimate their backgrounds.
Their very specific and concrete needs are stretching the limits
of perturbative QCD technology, as I mentioned, both for
calculating complicated processes at the level of quarks and gluons
and for turning these calculations into quantitative predictions
for practical observables (hadronization and jet algorithms).

There is also another field of application of QCD to high energy
physics
with quite a different flavor [\bur ].
One often attempts
to extract information about weak interaction
parameters (\eg \ Kobayashi-Maskawa angles and CP violation parameters)
from measurements of processes involving hadrons.
In the quantitative interpretation of such
measurements, uncertainty in estimates of hadronic matrix
elements of known operators is frequently the limiting factor.
It is quite embarrassing that we still don't have a really solid
understanding of the origin of the $\Delta I~=~ 1/2$ rule.
On a brighter note, in the past few years simple but profound
insights regarding the
universality of heavy quark couplings have blossomed into a richly detailed
and useful phenomenology [\wise ].

These applications reflect the maturity of QCD.
It is taking its place
beside its venerable
sister, QED, as a model of respectability and service.
The applications mentioned in the previous two paragraphs are rather
precisely parallel to the inclusion of QED radiative corrections
in high-energy processes and the use of QED to calculate atomic properties
for atomic parity-violation experiments, respectively.

$\bullet$ {\it  For nuclear physics}:  In principle, of
course, QCD provides the logical foundation for nuclear
physics in the same sense the QED provides the logical foundation
for chemistry.  In fact I think this analogy is quite appropriate,
because in both cases there is a wide gap
between principle and practice, and for similar reasons.
Chemistry will presumably always
use its own
semi-phenomenological language and methods
(valence bonds, molecular orbitals, ...),
only loosely connected to
fundamental concepts of QED, because the questions of most interest
to chemistry
involve delicate balances of
effects at energy scales of small fractions of
an electron volt, whereas the simple
fundamental interactions are characterized
by scales of several electron volts.  Similarly nuclear physics will
use its own semi-phenomenological language and methods (shell and liquid
drop models, pairing correlations, ... ) only loosely connected with
fundamental concepts of QCD.  This is because the energy scales of interest
to nuclear physics are an Mev or less, whereas the natural scale for
the fundamental interactions of QCD are at least a hundred times larger.
It is way too much
to expect .1\% calculational control of QCD
in the forseeable future as a practical matter.  On the other hand
it does not seem unreasonable to hope that major {\it qualitative\/}
aspects of nuclear phenomenology involving large energy scales, such
as the existence of the hard core, the saturation of nuclear forces,
and the rough magnitude and
distance dependence of the tensor interaction,
from fundamental principles.

Perhaps more glamorous is the prospect of exploring nuclear physics
in extreme or unusual circumstances, where the fundamental interactions
are less thoroughly balanced and masked.  Such possibilities arise
for example
when one considers baryonic matter with high strangeness content, at
high density and/or temperature, or when equilibrium is violently
disturbed
as in relativistic heavy ion collisions.  I shall say more about these
various possibilities below.

$\bullet$ {\it For astrophysics}:  Extreme nuclear
densities are reached in supernova collapse and in neutron star cores.
There are some very interesting QCD-based speculations for what might
happen under these conditions, as I shall discuss momentarily.

$\bullet$ {\it For cosmology}: Big bang cosmology forces you to consider
the behavior of QCD at extreme temperatures, since it leads one to expect
that a QCD plasma of quarks, antiquarks
and gluons filled the Universe and dominated
its matter content for at temperatures above $\sim$ 100 MeV.  These
occurred during the first 1/100 of a second or so after the Big Bang.
That might seem like a short interval of time, but of course it is
larger by many orders of magnitude than the mean free time between
collisions at that temperature, so that from the point of view of the
quarks and gluons things were developing at a very leisurely pace and
(in the absence of hysteresis, \ie\ supercooling) thermal equilibrium
should be accurately maintained.  So for the foundations of Big Bang
cosmology it is important to understand the nature of
near-equilibrium QCD at high temperature and
essentially zero baryon number -- its equation of state, possible
phase transitions, {\it etc}.  (Except in very heterodox models of
baryogenesis, the
asymmetry between quarks and
antiquarks was a few parts in a trillion at
these early times, so the approximation of zero baryon number is
appropriate.)  We do have some pretty good theoretical ideas and
numerical simulation results on these questions, as I'll discuss
below.

$\bullet$ {\it For numerical experiments}: Finally,
there is the charming domain
numerical experiments.  Here you have great flexibility:
you can simulate
different numbers of massless quarks, you can use  SU(2) or
E$_8$ as your strong
interaction gauge group, and so forth.
Of course I'm too young to
remember it, but I've read that in
the 60's there was a graffito
that appeared in Berkeley and became popular,
which read
{\it Reality is a crutch}.  This is a profound truth [\bohr ].
In any case, a
deep and intellectually satisfying understanding of QCD must
include ability to anticipate
how its predictions change as you change its parameters; something
one can only explore ``experimentally'' by numerical simulation.
Numerical experiments also allow you to look at
the underlying dynamics
in simple and directs ways
that are not at all practical in real experiments: for example,
one may look at the statistics of fluctuations in the gluon field,
look for instantons,
(perhaps) study the fully chiral symmetric theory with either two
or three massless quarks ... .

$\bullet$ {\it For natural philosophy}: With the success of QCD
the ``radically conservative'' principle that the behavior of the
physical world is captured in universal mathematical laws is once
more triumphant.  QCD gives no encouragement to doubts about the
foundations of quantum theory.  It is firmly based on received
principles of quantum field theory.  Indeed, it
allows tests of these
principles, including the effects of virtual particles and the
detailed workings of renormalization theory, far more extensive and
in many ways more stringent than was possible before.  I think
there are interesting reasons not to be {\it entirely\/} satisfied
with this outcome, however, as I'll mention at the end.

\section{Hot QCD [\stat ]}

At temperatures
well above the QCD scale, QCD is expected to be
qualitatively different from what it is at T = 0. Confinement should no
longer be taking place, the chiral condensate should melt or ionize.
It is very interesting, of course, to ask, are
there sharp phase transitions associated with these phenomena and
if there are, what is their nature?

The Wilsonian renormalization group
technology based on the concepts of scale invariance and universality
gives an entry into this question, because it tells us that
 to get some insight into these questions
you don't have to analyze QCD itself but just a theory with the same
symmetries  and very long wavelength modes, something in the same
universality class.  Furthermore if we are interested in the
properties nearest second order phase transition we should look for
a
scale invariant theory in the same universality class.
Such theories are very hard to come by.  If we find one then there is a
good chance that we
have actually found a model that  describes important aspects
of
the behavior of QCD near the phase transition precisely,
because all theories in the same universality class
exhibit
the same behavior.
I want to emphasize that we are speaking of {\it precise\/}
predictions even though we
are dealing with a strongly interacting theory under conditions that sound
horribly
complicated.

Working out
this program produces candidate second order phase transitions for the
confinement/deconfinement phase transition in pure glue SU(2), which in
the universality class of the Ising model, but not pure glue SU(3).
For
chiral symmetry restoration we have a candidate second order phase
transition for two massless quarks, which is in the universality class
of a 4-component Heisenberg magnet,
but no candidate for three or more massless quarks.
Thus you arrive at seemingly bizarre, highly
nonintuitive predictions
that whether one has a second order phase
transition or a first order one, that is whether
the free energy is continuous (but not analytic) or discontinuous
at the transition
depends on the exact number of colors or the number of quarks.

Remarkably, the evidence from lattice simulations
is fully consistent with these ideas so far [\brown ].
That is for SU(2) you have a second order phase transition with at least
the rough
character of the phase transition of the Ising model, for color
theories with several massless quarks you have what appears to
be a second
order phase transition for two massless quarks but
a first order transition for three or more.  (Actually the simulations
are not yet good enough to convince a reasonable skeptic; but it is fair
to say that they are consistent with prior theoretical expectations, and
do show striking changes of behavior as the number of species is varied.)

If there is a second order phase transition, that is if
we do have a scale invariant theory and a continuous transition near
the critical point then one can make obtain precise predictions
for QCD simply by going to the library and looking up calculations
done by our brethren in
condensed matter.  It is quite arduous to
obtain predictions for the scale invariant theory, even for the
Heisenberg magnet model (that is, for the symmetric n-component
massless $(\phi^2)^2$ theory), because it occurs at a large
value of the coupling constant.
In a remarkable {\it tour de force\/}
Baker, Nickel and Meiron calculated 353
graphs for the wave function renormalization and 789 graphs for the
vertex renormalization in the Heisenberg magnet model (that is, the
symmetric n-componenet $\phi^4$ theory)
-- see Figures 11 and 12 -- estimated the higher orders using
semiclassical techniques, and used Borel summation techniques
on a suitably
transformed coupling constant.
At the end of the day you obtain,
from a theory whose coupling constant is of order 1 in sensible
units,
two
significant figures in predictions for the critical exponents
characterizing the behavior near the phase transition.  So, for
instance, the specific heat is supposed to go like this:
$$
C(T) \sim A_\pm |t|^{-\alpha} + {\rm less\ singular}
$$
for $t \equiv \vert {T-T_c \over T_c} \vert$ with coefficient $A_+$
above and $A_-$ below $T_c$.  Here $\alpha = -0.19\pm .06$ and
$A_+/A_- = 1.9 \pm 0.2$; thus it has a cusp with a specific shape above
a continuous background.
This is one among a host of predictions.
There is, for example, a precise prediction
for the functional form of the condensate $\langle \bar q q \rangle$ as
a function of temperature and intrinsic quark mass near the transition
temperature [Figure 13].
This is the sort of thing that numerical simulations will presumably
be able to do well in the long run.  Present results
[Figure 14], while
encouraging qualitatively, are far too
crude for detailed comparison [\actual ].  Maybe by
the year 2000 we'll have it; the situation
doesn't look much worse than what we had in the
early days of testing for anomalous dimensions in deep inelastic
scattering.

There is a lot more to say about these
phase transitions, but perhaps these highlights give you a flavor
of what's possible.  I think it is
quite a remarkable thing that you can get precise predictions for
the behavior of QCD,
in these specific circumstances, at long distances and
high temperatures.

\section{Equation of state; QCD astrophysics}

Now I would like to discuss a specific problem in astrophysics
where better understanding of QCD would be very valuable.
Kaplan and Nelson [30] proposed, and Politzer and
Wise [31] have analyzed more quantitatively, the possibility that in dense
nuclear matter K$^-$ mesons condense.  Formally, this means that the
expectation value $\langle \bar u \gamma_5 s \rangle ~\neq~ 0$.
Its existence would mean that portions of ``neutron'' stars with
such high densities would in fact contain protons together with
a background Bose condensate of K$^-$ mesons.
The physical mechanism underlying the possibility of K$^-$ condensation
is simply that there is a strong attractive interaction between
these mesons and nucleons, which lowers their effective mass$^2$
-- perhaps
through zero --
in a dense medium.
These authors concluded that the possibility of such condensation is
not precluded for baryon number
densities roughly three times the usual
density of nuclear matter or more.

In a very intriguing paper Bethe and Brown [\bb ]
propose that this effect softens the nuclear equation of state, and
makes cold neutron stars with masses greater than 1.5 solar masses
unstable to collapse.  If true, this implies
that most progenitor stars with masses
greater than about 18 solar masses wind up as black holes after a
supernova explosion and stars with mass greater than 30 solar masses
go directly into a black hole.  This potentially solves a mystery in
the skies, the mystery of where is the star in Cassiopeia A?  There
was a supernova  explosion
there in historical times, somewhere in the period
1659-1675, as one infers from the size and motion of the
remnant ejecta.
Cassiopeia A it is one of the strongest radio
sources in the sky today, but there is no pulsar or
thermal x-ray source
in
the neighborhood, as one might expect to see for a remnant neutron
star.  This is actually an instance of a larger puzzle: after an
exhaustive
study of radio, optical, x-ray and $\gamma$-ray surveys Helfand
and Becker
[\helf ] came to the startling conclusion that nearly half the supernovae
in the galaxy leave no observable remnant.

Bethe and Brown suggest that in most of these cases
there is no neutron star
because
the progenitor
collapsed into a black hole following the supernova event.
They point out that there is a developing crisis with
supernova 1987a, which hasn't yet revealed any sign of
a neutron star in the middle, and they
suggest that in that case too there is a black hole.
If they're correct there
shouldn't be any neutron stars heavier than about 1.5 solar masses.

So there is a test of the theory of K-condensation, or at least this
consequence of K-condensation, there shouldn't be
any neutron stars with masses above about 1.5 solar masses.
Presently only the Vela pulsar challenges this bound; it
is reported to be 1.7 solar masses but there are
large uncertainties
in the measurement.

In any case I think its a fascinating possibility and
a great challenge that
considerations from
QCD potentially play a crucial role in astrophysics.  If
we could do better calculations, and
reliably decide under what circumstances K
condensation
and whether it or any other effect drastically
softens the nuclear equation of state at high density
we would be
performing a real service for astrophysics and nuclear physics.

\section{Strange matters}

A fascinating possibility,
suggested by Witten and others [\wit ] in the early 80's and analyzed in some
depth by Jaffe and others [\far ], is the possibility that there are compact
stable or metastable states
of finite baryon number carrying large strangeness,  and with much
higher density than ordinary nuclear matter.
 It is an outstanding challenge in QCD to
decide whether such states in fact exist, and to calculate their
properties.  The simplest candidate is the so-called H particle,
which has the quantum numbers of the (uuddss) 6-quark configuration.
Bag model estimates are formally consistent with the possibility that
this would actually be stable; within the uncertainties it might
also be below the $\Lambda~\Lambda$ threshold and thus unstable
only to weak decays, or above this threshold and therefore presumably
highly unstable.  There is an active effort to find this particle.
It clearly would be highly desirable to have a reliable
(accurate, though not necessarily extremely precise)
theoretical
estimate for its mass; to my mind it is something of a scandal that
lattice gauge theory has not produced one for us.

Less radical but perhaps more sure-fire, and
still extremely interesting, are {\it loosely\/} bound nuclear
molecules, some of which are predicted to be stable against
strong decay, but of course not against weak [\schaf ].
The spectroscopy of such states might provide
a testing ground for low-energy QCD which is complementary to, and
perhaps fresher than, conventional nuclear physics.

\section{Heavy ion collisions [\mul ]}

Parton cascade models, which although crude can hardly be
wrong qualitatively at high energies,
suggest that at proper time 1.5 fermi over c
following the collision for
gold on gold at 100 GeV per nucleon, that is under the conditions
attainable at RHIC,
gluons attain
a effective temperature of 325 MeV.
At LHC the attainable temperature is estimated to be
at least twice as large.
The quarks are found to be in
kinetic equilibrium, that is their kinetic energies
are characterized by a Boltzmann
distribution with the same kind of temperature, but do not reach
chemical equilibrium.
There aren't quite enough quarks and anti-quarks for the
given
temperature.

In any case, the estimated characteristic
temperatures are well above the
nominal QCD phase transition temperature, which
is probably not far from
150 MeV according to lattice simulations.
So the coming heavy ion collisions will
certainly take us into qualitatively new regimes, where the
fundamental constituents of hadronic matter and their
symmetries come into their own.

Unfortunately these regimes are only reached fleetingly
in the initial fireball, and
what one actually gets to observe is
what emerges from it after expansion and
cooling, a final state containing
many hundreds or even thousands of particles.
Under these conditions there is almost limitless potential
for uninspiring measurements that have no crisp interpretation.
However, there are also
a few important
qualitative effects so distinctive that their signature
might be discernible even in this noisy environment.

First, there is vastly
greater entropy in a quark-gluon plasma
than in a pion plasma at
the same temperature, because there are 8 colored gluons that
come with two helicities, plus two or three relevant flavors
of colored quarks and antiquarks with two helicities,
as opposed to a measly 3 scalar pions.
Making some bold but not unreasonable
approximations to set up the hydrodynamics of the
expansion, one can work out the observable consequences of
radically altered equation of state at high temperature
(if the quark-gluon plasma forms).  Crudely speaking, it is that
there should be a higher phase-space density in particles boiling
off at the early stages, which are generally those with the
highest transverse momentum [\van ].
It would be quite gratifying, though hardly shocking,
to see that this
dramatic increase in entropy actually
occurs.

A second thing that {\it might\/} happen -- this is much less certain --
is that the
chiral condensate melts [\raja ].  The temperature is certainly high enough,
and
the only question is whether this degree of freedom equilibrates in time.
If the chiral condensate does melt, then as it cools it will re-solidify.
It is as if you heated a magnet past its Curie point and then cooled it
back down; it will spontaneously magnetize again -- but perhaps with
the spins all aligned in a different direction!
Now in QCD it is not quite true that all directions for the chiral
condensate are equivalent; that would be true if the light quark masses
were rigorously zero, but they are not.  Nevertheless in the initial stages
the difference in energy density between alignments in different directions
is much less than the ambient energy densities, or the energies which
correlate the {\it relative\/} alignment at different points in space.
So there will be quasi-stable configurations where the chiral condensate
is assigned in the same wrong direction over a large region of space.
Putting off for a moment the question whether it is plausible that
such regions actually form, let us consider how they would evolve once
formed.  They will relax coherently toward the correct direction,
emitting coherent waves of pions as they do so --
we have a sort of pion laser.  The
radiation will consist of many pions with a fixed ratio of charged to
neutral, and low relative moment.  One can work out the distribution of
the neutral to charged ratio from simple geometry; it is
$$
{\rm Prob.} ({\cal R} \leq \lambda ) ~=~ {1\over 2} \lambda^{-{1\over 2}}~,
$$
where
$$
{\cal R} ~\equiv~ {N_{\pi^0}\over N_{\pi^0} ~+~ N_{\pi^+} ~+~ N_{\pi^-}}~.
$$
This is of course markedly different from the Gaussian centered at $1/3$
one would expect for uncorrelated emission, and could be very distinctive
even for rather modest numbers of pions in an emission region (defined
by small relative momenta).

Do such regions form?  Near equilibrium, we are really just asking
whether the correlation length becomes large near the phase transition.
That's
the signature for large correlated regions.
It sounds promising at first hearing
that there is likely a second-order
chiral transition for massless quarks, since that is precisely the
situation where you expect long correlation lengths.  Unfortunately
the quark masses are not zero, and for this application they are
too large.  One measure of this is that the
effective pion mass at the nominal transition temperature
is comparable
to the temperature itself (rather than zero as it would be for
massless quarks), so that a correlation volume will only contain enough
energy to create a small number of pions.

An interesting
physical mechanism for producing large correlated regions, which
might or might not be a good idealization of what happens
in a heavy ion collision, is amplification after a quench.
A quench is
what happens if you suddenly change the temperature of
a system, removing the energy in highly excited modes but not
in low-lying modes which do not have time to equilibrate.  The
prototype quench is to plunge a hot iron bar suddenly into ice
water.  The free energy functional changes suddenly from being
the free energy functional appropriate to high temperature to the one
appropriate to  zero temperature or very low temperature, which
(if we have cooled through a phase transition temperature) may
lead to a change in symmetry.  In terms of
pion dynamics specifically we have the following.  The frequency of
oscillation of the pion field is given by a dispersion relation
$$
\omega^2~=~ k^2 ~ - ~\mu^2 ~ + 2\lambda v^2
$$
where $v$ is the expectation value of the magnitude of the chiral
condensate, $\lambda$ is the coupling and -$\mu^2$ the (negative)
bare mass$^2$.  In the ground state the two last terms
on the right cancel, and we have massless Nambu-Goldstone bosons.
(I am ignoring the quark masses at this point; they do not alter the
essence of the matter.)  However if we start with $v^2~=~0$, as at
the beginning of a quench, then the effective mass$^2$ is negative,
and the frequencies are imaginary indicating the possibility of
exponential amplification.  The amplification factor is larger,
the smaller is $k^2$ -- thus long-wavelength modes are the most
enhanced, as we desired to show.

Finally one expects copious production of strange quarks and
antiquarks at high temperatures, so that conditions are
favorable
for production of the strange matters discussed above, and it will
be important to search for them in the debris of heavy-ion collisions.

\chapter{Foundational Issues}

Now I'd like to discuss some outstanding foundational questions.
The purpose of this part is more to raise questions than to provide
answers.

\section{Does perturbation theory suffice?}

We have already touched on to some extent
on the question of how far can one push perturbation theory.  We argued,
following Mueller and others, that it is possible
to push it quite far as an asymptotic expansion for
large momenta.  What is the general situation?

It is very common for
people to talk loosely as if there were one thing called perturbation
theory and another thing called non-perturbative effects,
as if
non-perturbative effects were a world apart.
However it is also conventional wisdom -- although perhaps
{\it unconscious\/} conventional wisdom --
to think that perturbation theory in fact determines the theory
completely.  Indeed it is a very simple matter to read off the
Lagrangian from low-order graphs, and the Lagrangian is normally
supposed to determine the theory completely, even non-perturbatively.

This ``determination in principle'' is of course a very different
matter from claiming that the perturbation theory converges,
or  providing
a constructive procedure to go from it to the correct answer.
I am not aware that anyone has even made a really plausible conjecture
for how to do it, in the spirit of Borel resummation.
(Borel resummation
itself will not work, for reasons mentioned before.)

If one could find it, such a procedure would be informative
in many
ways.  For one thing, it would allow one at least to truly
{\it define\/} chiral gauge theories, and to regulate QCD
in a manner that is manifestly chirally symmetric.
We certainly
know how to
construct the perturbation series for such theories, which I
remind you include such interesting examples as
the Weinberg-Salam model and essentially all realistic unification
schemes.  It is less clear that we know how to regulate them
non-perturbatively (but see below).   It could also be instructive
for string theory, where one has very little besides the perturbation
expansion to go on, and the same problems arise in what appears
at least superficially to be a more severe form.

Now actually it is not quite true that the perturbation theory
defines QCD completely.  QCD has -- it is commonly believed --
one additional parameter, the famous angle $\theta$, whose effects
do not show up in any order of perturbation theory.  Nevertheless I
believe there is a visible
signal for the existence of such a parameter in
perturbation theory.  In analyzing the high order behaviors of the
coefficients in perturbative expansions, one finds that classical solutions
of the Euclidean field equations play a crucial role.  They induce
specific forms of the growth of high-order coefficients in the perturbative
expansion of all amplitudes, which can be inferred from the classical
solution.  Conversely, by comparing the high-order coefficients in the
perturbative expansion of various processes one can reconstruct the
classical solution.  In this way QCD instantons, to be specific, are
visible upon close scrutiny of perturbation theory.  Now whatever
resummation procedure one uses to take care of these contributions, it
must contain an ambiguity corresponding to the possibility
of introducing different values for the $\theta$ parameter.
Based on the standard formal analysis of the continuum theory [\call ],
I suspect
that one will find that the resummed series will not define amplitudes
which obey the cluster decomposition axiom unless one
adds together
weighted sums of graphs for perturbatively distinct processes in
defining the amplitudes, just as one must pass in the formal
continuum analysis
from n-vacua to $\theta$-vacua.

This is an
example of how supposedly non-perturbative effects such as
axial baryon number violation in QCD with massless quarks -- or
just plain baryon
number violation in SU(2)$\times$U(1) -- can and should be visible,
in principle, in perturbation theory.  It also brings home, I hope,
that
the problem of behavior of high orders and
resummation in perturbation theory
is definitely not an insipid technicality;
it encodes deep yet tangible physical phenomena.  It natural to
ask: are there
specific perturbative signatures of confinement? of chiral
symmetry breaking?

\section{Chiral symmetry on the lattice?}

There have been promising suggestions recently for how to put fermions on
a lattice, while respecting chiral symmetry [\nara ].  Can these methods be
made practical, and allow one to accomplish what one normally hopes
to
accomplish using lattice techniques?  Specifically: can one compute
a strong-coupling spectrum that does a reasonable job on the
pseudoscalar mesons -- both the octet and the $\eta^\prime$? Can one use
the scheme together with importance sampling, for
practical numerical work?

\section{What are the string rules telling us?}

I mentioned before that there are new methods for doing QCD
perturbation
theory, which seem to be remarkably efficient for some purposes,
whose formulation has been largely guided by string theory techniques.
The central result of the analysis is a set of rules, different from
the Feynman rules, for graphs.  It must be emphasized that at the
end of the day there is no reference to strings; the rules are
rules for calculating quantum field theory processes.

To my knowledge there is no
straightforward derivation of the rules that gives complete insight
into
their origin.  (For some partial successes, see [\lam , \strassler ].)
It is quite suggestive that rules can be formulated
in a way that involves following color flows and spin flows through
the diagrams [\lam ].  So we are led to ask, concretely:
is there a first-quantized action for particles, including internal
degrees of freedom,
that naturally leads to the string rules [\shatashvili ]?
One could conceivably hope to abstract from the relationship between
the properly formulated particle theory and its field theory,
insight into the poorly understood relationship
between strings and their field theory.

\section{How does God do it?}

Finally I would like to mention a question that in various forms has
bothered a few physicists [\deu ], and bothers me too.  It is that in every
approach I know to defining QCD, one must appeal to completed infinities.
Specifically, in a lattice approach one must say that the real amplitude
is computed by calculating on a lattice with a given spacing and then
taking the limit as the spacing goes to zero; in the (incompletely
formulated) approach of resumming perturbation series one
will almost certainly have
to compute every term in an infinite series,
perform analytic continuations,
... .  These tasks are
extremely computationally intensive: in fact, they
call for an infinite amount of computation.  Yet somehow
the good Lord
manages to get the results
quite effortlessly in small amounts of time.
Does this indicate that the mathematical notion of computability is
fundamentally
different from the physical one?  Can one imagine physical systems with
more computational power than Turing machines?  Are non-trivial
quantum field theories in a finite volume examples of such systems?
Or does a careful consideration of limitations on the actual observables
show that most of this
apparent computational power is never really called forth?

One aspect of these questions, I suspect the deepest,
is whether the existence of an infinite number of
degrees of freedom in any finite volume is physically
acceptable.  Closely related problems
arise in a slightly more conventional physical context, in the
physics of black holes.  When one takes the ground state
(or any reasonable excited state) of any
reasonable quantum field theory, and traces over the degrees of freedom
inside a specified spatial region, the entropy of the resulting density
matrix is infinite [\holz , \soe ].  In the presence of a
black hole one is forced to take just such a
trace, over the inaccessible degrees of freedom inside the horizon.
(In fact the difference between flat space and a black hole
in this regard is probably very slight, because the crucial correlations
are those between nearby points just inside and just outside
the horizon, and for a large black hole the curvature there
is small.)  This
is an infinite correction to the Bekenstein-Hawking entropy, which
presumably
is physically undesirable [\holz ].  Since the source of this infinity is
an ultraviolet catastrophe, it might be relieved in string theory; but
the calculations necessary to reach a definite conclusion
in this matter seem quite
challenging [\suss , \lars ].

So perhaps there are signs that the deep principles
of quantum field theory that QCD embodies so perfectly
are
not the last word about ultimate
reality.

\bigskip

\bigskip

There is much more to say about each of the topics
I have mentioned,
and there are many important,
vital topics in QCD that I have not been able to
mention at all, for which I apologize.
My summary of the status of QCD was given at the beginning, and I
hope that now you find it more plausible.
\refout
\vfil
\eject
{\bf FIGURE CAPTIONS:}
\bigskip
\bigskip
\noindent
Figure 1: Determinations of the effective coupling $\alpha_s (Q^2)$,
compared with QCD predictions (updated from Bethke, Ref. [2].  I
thank Professor Bethke for supplying this Figure, and also the
following two.)
\bigskip
\noindent
Figure 2:  Summary of theoretical foundations and uncertainties for
the determinations in Figure 1.
\bigskip
\noindent
Figure 3:  Tabular form of Figure 1.
\bigskip
\noindent
Figure 4: Cartoons referred to in the analysis of current products.
\bigskip
\noindent
Figure 5: The variation of the structure function $F_3(x, Q^2)$ with
\Q.  QCD predicts an $x$-dependent logarithmic decrease, reflecting
the softening of the effective quark distribution by gluon radiation.
\bigskip
\noindent
Figure 6:  Relative percentage of 3-jet as opposed to 2-jet events in
$e^+e^-$ annihilation.  This reflects directly the strength of the effective
coupling, since the third jet arises from this coupling.
\bigskip
\noindent
Figure 7:  The angular distribution of jet in pp collisions, if they
are supposed to arise from exchange of quarta with different spins,
or from QCD.
\bigskip
\noindent
Figure 8:  The actual data, showing a good fit to QCD.
\bigskip
\noindent
Figure 9:  Data regarding a coherence effect in jet physics,
explained in the text.
\bigskip
\noindent
Figure 10:  QCD as a service subject:  what it can offer other
fields, and conversely what questions in QCD are stimulated
by other fields.
\bigskip
\noindent
Figures 11, 12: Some of the graphs calculated by Baker, Meiron, and
Nickel to obtain accurate critical exponents.
\bigskip
\noindent
Figure 13: An example of the information that can be extracted from
analysis of critical behavior in 2-light flavor QCD.
\bigskip
\noindent
Figure 14: Numerical data, of the kind ultimately will be confronted
with Figure 13.

\bye